\documentclass[5p,twocolumn,times,number]{elsarticle}

\usepackage{graphicx}
\usepackage{amsmath}

\begin{document}

\begin{frontmatter}

\title{Measurement of neutron detection efficiency between 22 and 
  174 MeV using two different kinds of Pb-Scintillating fiber sampling 
  calorimeters}

\author[add1]{M.~Anelli}
\author[add1]{S.~Bertolucci}
\author[add2,add3]{C.~Bini} 
\author[add5]{P.~Branchini}
\author[add1]{G.~Corradi}
\author[add1]{C.~Curceanu}
\author[add2,add3]{G.~De~Zorzi} 
\author[add2,add3]{A.~Di~Domenico}
\author[add4,add5]{ B.~Di~Micco}
\author[add6]{A.~Ferrari}
\author[add2,add3]{S.~Fiore}
\author[add2,add3]{P.~Gauzzi}
\author[add1]{S.~Giovannella}
\author[add1]{F.~Happacher}
\author[add1,add7]{M.~Iliescu}
\author[add1]{A.~Luc\`a}
\author[add1]{M.~Martini}
\author[add1]{S.~Miscetti\corref{cor}}
\ead{stefano.miscetti@lnf.infn.it}
\author[add4,add5]{F.~Nguyen}
\author[add5]{A.~Passeri}
\author[add8]{A.~Prokofiev}
\author[add1]{I.~Sarra}
\author[add1]{B.~Sciascia}
\author[add1]{F.~Sirghi}
\author[add1]{D.~Tagnani}

\cortext[cor]{Corresponding author}

\address[add1]{Laboratori Nazionali di Frascati, INFN, Italy,}
\address[add2]{Dipartimento di Fisica dell'Universit\`a ``La Sapienza'', 
  Roma, Italy}
\address[add3]{INFN Sezione di Roma, Roma, Italy}
\address[add4]{Dipartimento di Fisica dell'Universit\`a ``Roma Tre'', 
  Roma, Italy}
\address[add5]{INFN Sezione di Roma Tre, Roma, Italy}
\address[add6]{Fondazione CNAO, Milano, Italy,}
\address[add7]{IFIN-HH, Bucharest, Romania,}
\address[add8]{The Svedberg Laboratory, Uppsala University, Sweden.}

\begin{abstract}
We exposed a prototype of the lead-scintillating fiber 
KLOE calorimeter to neutron beam of 21, 46 and 174 MeV at
The Svedberg Laboratory, Uppsala, to study its neutron 
detection efficiency. This has been found larger than what expected 
considering the scintillator thickness of the prototype.
We show preliminary measurement carried out with a different prototype
with a larger lead/fiber ratio,
which proves the relevance of passive material to neutron detection 
efficiency in this kind of calorimeters.
\end{abstract}


\end{frontmatter}


Detection of neutrons with energies from a few to hundreds MeV is usually 
performed with organic scintillators, where the elastic scattering of 
neutrons on hydrogen atoms produces a visible response from recoil protons. 
Typical efficiency is $\approx 1\%$ per cm of scintillator thickness.
The insertion of an organic scintillator in a high Z material, with a 
sizable cross section for elastic and inelastic neutron interactions, 
could originate a large production of secondary particles and consequently 
increase the detection efficiency.

The fine sampling lead-scintillating fiber KLOE calorimeter \cite{NIMcalo} 
has been primarily designed to detect low energy photons.
During a study of kaon interactions in KLOE, a high efficiency for low
energy neutrons was observed and then confirmed by the experiment simulation.
To understand the underlying physical mechanisms which produce this 
difference, we planned a set of test beams to expose the calorimeter
to dedicated neutron beams and we carried out a full simulation 
of the detector.


The KLOE calorimeter prototype used in this measurement is made of 
$\sim$ 200 layers of 1~mm diameter blue scintillating fibers, glued 
inside grooved lead layers of 0.5 mm thickness. The final structure 
has a fiber:lead:glue volume ratio of 48:42:10 resulting in a density 
of $\sim$ 5~g/cm$^3$. 
The total external dimensions are $( 13 \times 24 \times 65 )$~cm$^3$, 
where the second value is the calorimeter depth and the third one is the 
fiber length. The calorimeter is readout at both fiber ends to reconstruct 
this coordinate by time difference. 
The readout is organized in four planes in depth and three  
columns along the horizontal coordinate, originating cells of 
$( 4.2 \times 4.2)$ cm$^2$.  
Larger readout elements are used in the rear part of the calorimeter. 
Each element is coupled to standard photomultipliers, PM's, through 
light guides.
The PM signals are split to form the trigger with the coincidence
of their discriminated analog sum for each side.

A reference counter was built with 
a 5~cm thick bulk of NE110 organic scintillator, of transversal 
dimensions $(10 \times 20)$~cm$^2$, by coupling it at the two ends 
to two EMI9814 PM's.
To trigger on the scintillator, the PM signals were discriminated 
and an overlap coincidence was formed.

We have run our experiment at ``The Svedberg Laboratory'' (TSL) 
neutron beam facility~\cite{TSL}, performing different test 
beams with high energy neutrons (174~MeV) in October 2006 and 2008 
and low energy neutrons (21 and 46~MeV) in June 2007.
The neutron energy spectrum is dominated by a peak at few MeV below 
the primary proton energy and a long tail down to thermal neutrons. 
Low intensity neutron beams of few kHz/cm$^2$ has been required to 
minimise the probability of double neutron counting. 
The neutron rate, $R_{n}$, has been measured by an Ionization-Chamber 
Monitor, ICM, with an absolute accuracy of 10\% (20\%) at high (low) 
energy.
For a given trigger threshold, assuming full beam acceptance and no 
background, the efficiency of the detector to the overall neutron spectrum 
has been determined according to the formula:
$\varepsilon = R_{\rm DAQ} / (R_n \cdot F_{\rm live})$,
where $R_{\rm DAQ}$ is the acquired rate for the detector and
$F_{\rm live}$ is the fraction of DAQ live time.


A detailed simulation of the calorimeter structure and of the experimental 
beam line has been done using FLUKA \cite{FLUKA}, which 
computes the energy deposits in the scintillating fibers, taking into 
account the signal saturation due to the Birks' law. 
A data-MC comparison of the neutron time of flight (ToF) distribution 
indicates a contamination of events coming from the area surrounding 
the collimator (halo). 
Its contamination is obtained by fitting the ToF data distribution with 
the expected signal shape from MC and the halo contribution.
The halo shape 
is obtained both from the outer calorimeter cells and from dedicated runs 
with the calorimeter out of the beam line.
The halo contamination is higher for 
events with low cell multiplicity, with an overall contribution of 
$\sim 30\%$ in the high energy runs.
This fraction is smaller for low energy neutron beams, in agreement with 
the measurements carried on with a fission monitor counter
used for the absolute calibration of the beam flux, which provide a 
correction factor of $0.80\pm 0.13$.


\begin{figure}[!t]
  \centering
  \vspace{-0.3cm}
  \includegraphics[width=0.50\textwidth,height=0.5\textwidth]{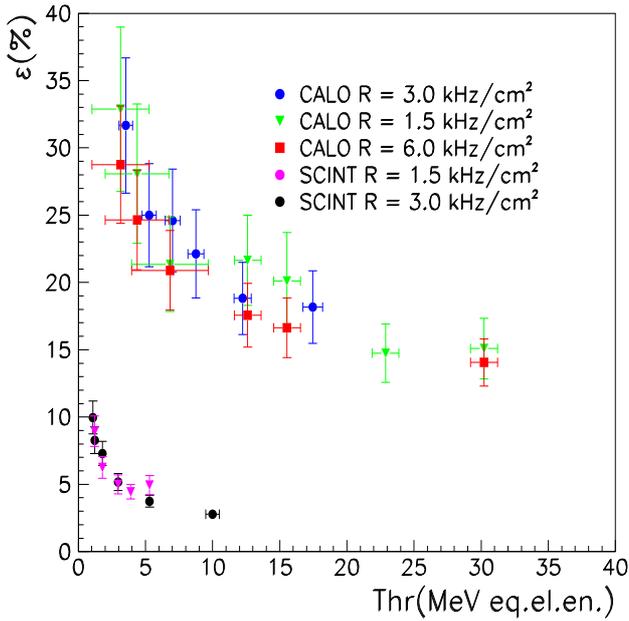}
  \vspace{-0.3cm}
  \caption{Dependence of $\epsilon_{calo}$ on the applied trigger 
    threshold for run at 174~MeV.
    The scintillator efficiency obtained with the same data, after
    scaling to the calorimeter equivalent thickness, is also reported.}      
  \label{effi_calo174}
\end{figure}

To evaluate the neutron detection efficiency, each cell is calibrated 
with minimum ionising particles (MIPs).
The ratios between data energy distributions at different threshold 
levels have been used to determine the cut-off introduced by the trigger.
After applying the beam halo correction, the calorimeter neutron detection 
efficiency ranges between 30\% at high energy (Fig.~\ref{effi_calo174}) 
and 50\% for low energy runs. 
The errors on vertical scale are dominated by halo subtraction and
absolute neutron flux, while on the horizontal scale a conservative 
error has been assigned.
For comparison, the efficiency of the 5~cm thick NE110 scintillator 
ranges from 4\% to 10\% for values of the trigger threshold below 5 MeV 
of electron equivalent energy, in good agreement with the available 
measurements in literature. This indicates that the measured calorimeter 
efficiency is sizeably enhanced with respect to the expected 8$-$10\% 
based on the amount of scintillator only. 


A second lead scintillating fiber calorimeter prototype, proto-2, has 
been tested with 174 MeV neutrons at TSL in October 2008. This prototype, 
32 cm in length and $(7.5 \times 7.5)$ cm$^2$ in cross section, has a 
different structure, with fibers at the vertices of squares, and a 
resulting structure with lower fiber/absorber ratio than KLOE:
fibers\,/\,total volume is 19.5\%.
At each module end, the fibers are grouped together in two bundles that 
are directly connected to PM's.
The trigger requires an analog sum of the two signals from each 
module end larger than a given threshold $V_{th}$. 

\begin{figure}[!t]
  \centering
  \vspace{-0.3cm}
  \includegraphics[width=0.50\textwidth,height=0.4\textwidth]{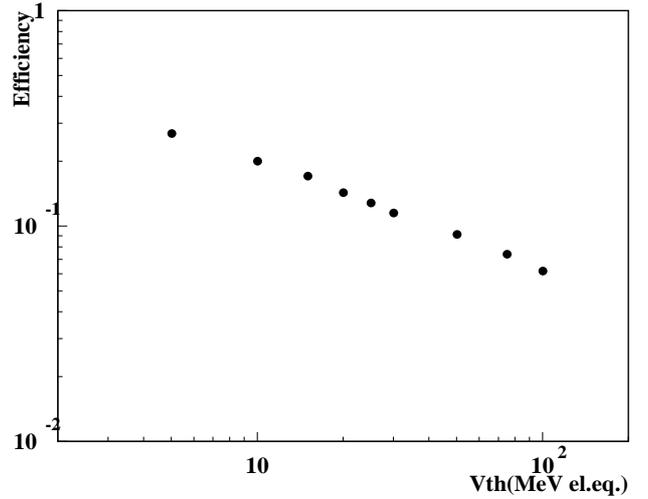}
  \vspace{-0.4cm}
  \caption{Efficiency of proto-2 detector as a function of the 
    threshold.}
  \label{effmev}
\end{figure}

The efficiency has been evaluated at different $V_{th}$ values with the
same technique described before. The contribution of the halo neutrons,
of the order of 10\%, is subtracted to the rate. 
The trigger threshold $V_{th}$ is converted in units of MeV equivalent 
energy by using the response of the detector to minimum ionizing particles 
and the quantity (e/MIP) given in Ref.\ \cite{Wigmans} for calorimeters 
having the same fiber/absorber ratio. 
Fig.~\ref{effmev} show the preliminary neutron efficiencies of proto-2 
as a function of $V_{th}$.
The neutron detection efficiency is comparable 
with that of the KLOE-like calorimeter, which has a larger fiber/absorber 
ratio but similar lead amount. This shows that the fraction (or the 
absolute quantity) of passive absorber plays an important role. If we 
compare the efficiency parametrizing it per unit of scintillator thickness
we get 3\%/cm and 13\%/cm for KLOE prototype and proto-2 respectively, 
which indicates a large efficiency enhancement for denser calorimeters.

\end{document}